\documentclass[aps,pra,twocolumn,floatfix,superscriptaddress,longbibliography]{revtex4-1}

\usepackage{xcolor}
\usepackage{appendix}
\usepackage{graphicx,times}
\usepackage{amstext}
\usepackage{amsmath}            %serve per le subequazioni
\usepackage{amssymb}            %serve per il simbolo "marchio registrato", \circledR
\usepackage{graphicx}           %serve per le figure eps, ps etcyy
\usepackage{latexsym}
\usepackage{bm}
\usepackage{color}
\usepackage{etoolbox}
\usepackage{breqn}
\usepackage[FIGTOPCAP,raggedright,nooneline]{subfigure}

\makeatletter
\let\cat@comma@active\@empty
\makeatother

\usepackage{url}
\usepackage[colorlinks]{hyperref}
\hypersetup{%
    plainpages=true,
    breaklinks=true,%not default in dvips mode, so we must specify
    hypertexnames=false,%not ideal, but needed when pagenums duplicate (`i' vs. `1')
    pageanchor=true,
    colorlinks=true,
    linkcolor={blue},
    citecolor={red},
    urlcolor={blue},
    anchorcolor={black}
}

\newcommand{\beq}{\begin{eqnarray}}
\newcommand{\eeq}{\end{eqnarray}}

\def\<{\langle}
\def\>{\rangle}

\def\<{\langle}
\def\>{\rangle}

\def \info#1{}

\def \info#1{}

\begin{document}

\title{Collectively induced exceptional points of quantum emitters coupled to nanoparticle surface plasmons}

\author{Po-Chen Kuo}
\affiliation{Department of Physics, National Cheng Kung University, Tainan 70101, Taiwan}
\author{Neill Lambert}
\affiliation{Theoretical Quantum Physics Laboratory, RIKEN Cluster for Pioneering Research, Wako-shi, Saitama 351-0198, Japan}
\author{Adam Miranowicz}
\affiliation{Theoretical Quantum Physics Laboratory, RIKEN Cluster for Pioneering Research, Wako-shi, Saitama 351-0198, Japan}
\affiliation{Faculty of Physics, Adam Mickiewicz University, 61-614 Pozna$\acute{\rm{n}}$, Poland}
\author{Hong-Bin Chen}
\email{hongbinchen@gs.ncku.edu.tw}
\affiliation{Department of Engineering Science, National Cheng Kung University, Tainan 70101, Taiwan.}
\affiliation{Center for Quantum Frontiers of Research \& Technology, NCKU, Tainan 70101, Taiwan}
\author{Guang-Yin Chen}
\email{gychen@phys.nchu.edu.tw}
\affiliation{Department of Physics, National Chung Hsing University, Taichung 40227, Taiwan.}
\author{Yueh-Nan Chen}
\email{yuehnan@mail.ncku.edu.tw}
\affiliation{Department of Physics, National Cheng Kung University, Tainan 70101, Taiwan}
\affiliation{Center for Quantum Frontiers of Research \& Technology, NCKU, Tainan 70101, Taiwan}
\author{Franco Nori}
\affiliation{Theoretical Quantum Physics Laboratory, RIKEN Cluster for Pioneering Research, Wako-shi, Saitama 351-0198, Japan}
\affiliation
{Department of Physics, The University of Michigan, Ann Arbor, Michigan 48109-1040, USA}
\date{\today}

\begin{abstract}
Exceptional points, resulting from non-Hermitian degeneracies, have the potential to enhance the capabilities of quantum sensing. Thus, finding exceptional points in different quantum systems is vital for developing such future sensing devices. Taking advantage of the enhanced light-matter interactions in a confined volume on a metal nanoparticle surface, here we theoretically demonstrate the existence of exceptional points in a system consisting of quantum emitters coupled to a metal nanoparticle of subwavelength scale. By using an analytical quantum electrodynamics approach, exceptional points are manifested as a result of a strong coupling effect and observable in a drastic splitting of originally coalescent eigenenergies. Furthermore, we show that exceptional points can also occur when a number of quantum emitters is collectively coupled to the dipole mode of localized surface plasmons. Such a quantum collective effect not only relaxes the strong-coupling requirement for an individual emitter, but also results in a more stable generation of the exceptional points. Furthermore, we point out that the exceptional points can be explicitly revealed in the power spectra. A generalized signal-to-noise ratio, accounting for both the frequency splitting in the power spectrum and the system's dissipation, shows clearly that a collection of quantum emitters coupled to a nanoparticle provides a better performance of detecting exceptional points, compared to that of a single quantum emitter.
\end{abstract}

\maketitle

%\pacs{03.65.Ud, 42.50.Dv, 03.65.Yz, 73.23.-b}

%%%%%%%%%%%%%%%%%%%%%%%%%%%%%%%%%%%%%%%%%%%%%%%%%%%%%%%%%%%%%%%%%%%%%%%%%
\section{INTRODUCTION}
The rapid development of quantum technologies have triggered intense interest in the potential of quantum sensors \cite{Degen2017,Pirandola2018}. Without considering energy loss or gain, one only needs a hermitian Hamiltonian to describe an energy-conserving system, where a diabolical point (DP) \cite{Seyranian2005}, containing degenerate eigenenergies with different corresponding eigenvectors, may be found. In realistic systems, however, one must consider the energy exchange process with an environment \cite{Weiss2012}, which in some situations can be described by an effective non-Hermitian Hamiltonian. 

An intriguing property of non-Hermitian Hamiltonians is that the degeneracy of eigenenergies can occur alongside the coalescence of the corresponding eigenstates, i.e., the occurrence of exceptional points (EPs) \cite{Ganainy2018,Ozdemir2019}. Owing to different mathematical properties of DPs and EPs, when the system is subject to a perturbation, the resulting energy splitting of a spectrum is shown to follow a square-root dependence on the perturbation at an EP, instead of being linearly proportional to the perturbation, as occurs at a DP \cite{Ozdemir2019,Demange2012,Lau2018}. In other words, the energy splitting of the spectrum at an EP may have an extremely sensitive dependence on the parametric change caused even by a small perturbation. This is why the splitting near an EP may be exploited for ultrasensitive sensing \cite{Ozdemir2019,Hodaei2017,Weijian2017}.

We note, however, that the true applicability and usefulness of EP sensing depend on the details of how the parametric change is measured \cite{Lau2018,Langbein2018}. In any case, finding practically useful EPs in physically accessible systems \cite{Peng2014,Arkhipov2019,L2017} and parameter regimes is still an open problem \cite{Mirieaar2019,Fabrizio2019,Ievgen2019}, and a range of candidates have been studied, such as parity-time-symmetric systems \cite{Christian2010,Regensburger2012,Peng2014,ZhongPeng2016,Hassan2015,Jing2014,Jing2015,Zhang2015,Quijandr2018}, coupled atom-cavity systems \cite{Choi2010}, microcavities \cite{Peng2014,ZhongPeng2016,Hassan2015,Lee2009,Wiersig2016,Zhu2010}, microwave cavities \cite{Dembowski2001,Dembowski2004,Dietz2007,Liu2017}, acoustic systems \cite{Ding2016}, photonic lattices \cite{Alfassi2011,Regensburger2012}, photonic crystal slabs \cite{Zhen2015}, exciton-polariton billiards \cite{Gao2015}, plasmonic nanoresonators \cite{Kodigala2016}, ring resonator \cite{Sunada2017}, optical resonators \cite{Jing2017,Peng2014science,Zhang2018}, and topological arrangements \cite{Gao2015}.

However, the typical size of these systems possessing EPs is usually too large (of several hundred nanometers) to be utilized for sensing in some important applications. Nevertheless, for such a nanoscale, the relevant parameters, such as coupling strength between objects, are not easy to reach the requirement of forming an EP. Fortunately, when light is incident on a metal nanoparticle (MNP), local oscillations of electrons, known as localized surface plasmons (LSPs), can occur at a length scale much smaller than the wavelength of light \cite{Tame2013,Chikkaraddy2016,Pile2017,GuangYin2011}. This implies that by placing quantum emitters (QEs), such as biomolecules, near an MNP, the electromagnetic field outside the MNP becomes tightly localized around the metal surface, giving rise to possible strong couplings between the QEs and MNP \cite{Savasta2010,Delga2014,Zhou2016}. Under suitable dissipation conditions, the existence of EPs at a subwavelength scale becomes possible.

In this work, we first predict the existence of an EP in the case of an MNP coupled to a QE, which is generically described as a two-level system. The role of QE can be played by, e.g., a biochemical molecule or a quantum dot. Surprisingly, we find that EP can also occur when several QEs are collectively coupled to the dipole mode in the MNP. Such a quantum collective effect not only relaxes the strong-coupling requirement for the individual QE, but also results in more feasible conditions to generate an EP. Additionally, by implementing a photon detector near the QE-MNP system, we show that the observation of an EP , as well as the frequency splittings in power specrum, is experimentally accessible.

Moreover, we analyze how accurately an EP can be detected by comparing frequency splitting in theoretical eigenenergy spectra and the output power spectra. Note that the required accuracy of the observation of an EP is limited by dissipation. To specify the degree to which the occurrence of an EP is affected by dissipation, we propose to use the signal-to-noise ratio taking into account both frequency splitting and dissipation. From our analysis of the signal-to-noise ratio, we conclude that a collection of QEs coupled to an MNP provides a better performance of the detection of EPs compared to that of a single QE.

%------------------------------------------------------------------------------------------------

\section{Single quantum emitter coupled to the silver nanoparticle}\label{sec:model}

\begin{figure*}[th]
\centering
\includegraphics[width=1.8\columnwidth]{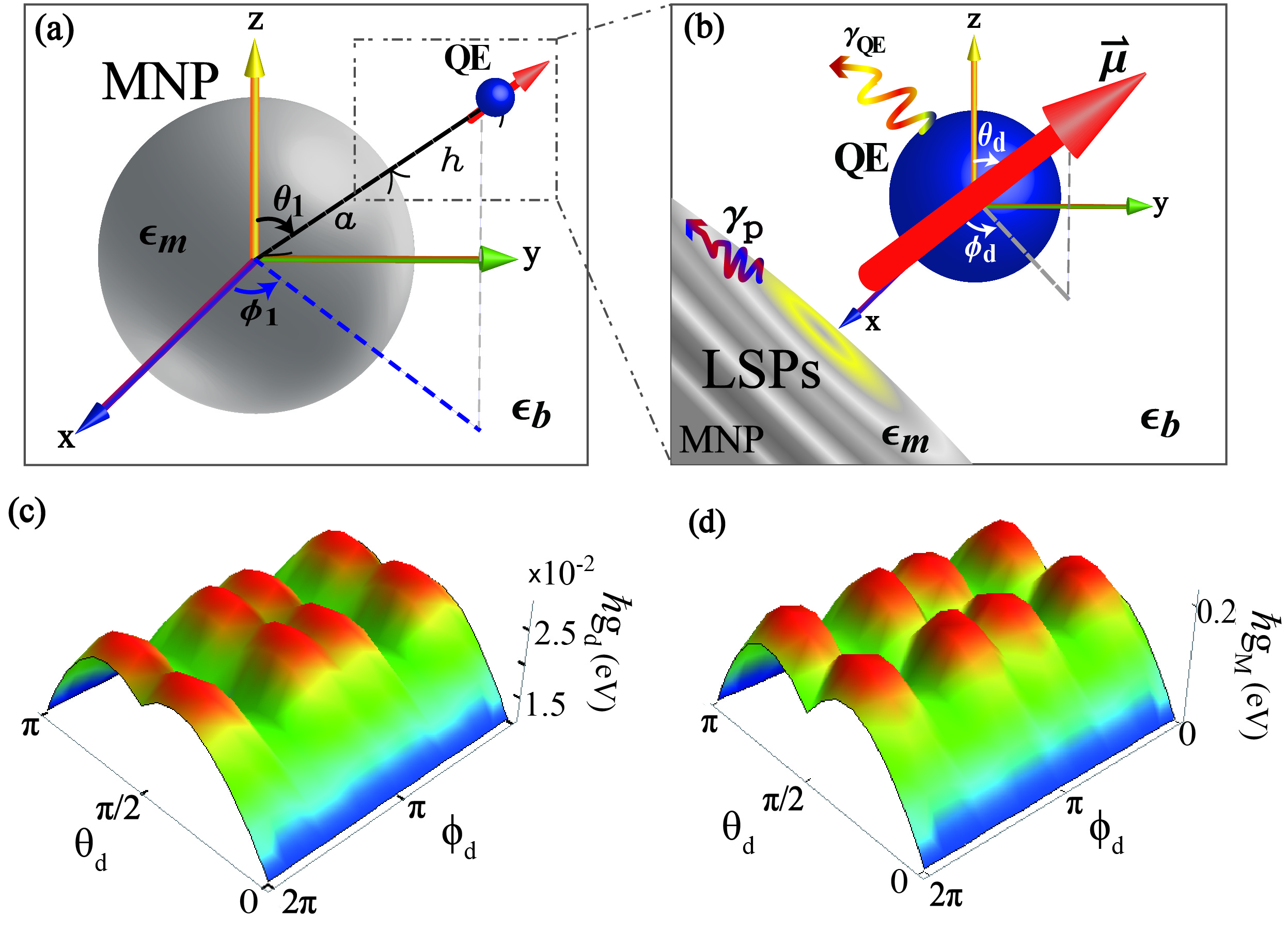}
%width=6.5cm,height=11cm%
\caption{(Color online) Schematic diagram of the QE-MNP model in spherical coordinates, and examples of the the strength of the coupling to the dipole mode and pseudomode with different dipole moment orientations. (a) A single quantum emitter embedded in a dielectric medium with permittivity $ \epsilon_{b} $ is in proximity to a silver metal nanoparticle at the position $ (r_{1},\theta_{1},\phi_{1}) $, where $r_{1}=a+h$. (b) Partial enlargement of (a): A quantum emitter with the dipole orientation $\vec{\mu}=\left(\mu_{r},\mu_{\theta},\mu_{\phi} \right)$ coupled to the localized surface plasmons with damping rate $ \gamma_{\rm{p}} $ as well as other decay channels, such as internal nonradiative decay and spontaneous decay into the dielectric material, with a total rate $\gamma_{\rm{QE}}  $. The strengths of the coupling to the dipole mode $\hbar g_{d}$ (c), and to the pseudomode $\hbar g_{M}$ (d), as functions of $ \theta_{\rm{d}} $ and $ \phi_{\rm{d}} $ with $ h =1 $ nm.} \label{Figure1}
\end{figure*}

In order to explore the possibility of using the emitter-plasmon system as a quantum sensor, we follow the formalism of Ref.~\cite{Delga2014}. Thus, we first consider a composite system embedded in a nondispersive, lossless dielectric medium with permittivity of $ \epsilon_{\rm{b}}=2.3 $, composed of a two-level QE close to the surface of a silver MNP with a distance $ h $, as depicted in Fig.~\ref{Figure1}(a). The MNP with a radius $ a=7 $ nm can be characterized by a Drude-type permittivity $ \epsilon_{\rm{m}}=\epsilon_{\infty}-\omega_{\rm{p}}^{2}/[\omega(\omega+i\gamma_{\rm{p}})] $ with $\epsilon_{\infty}=4.6 $, $\hbar\omega_{\rm{p}}=9 $ eV and the dissipation of silver $\hbar\gamma_{\rm{p}}=0.1$ eV. Here, the QE acts as a point-like dipole with the distance $ h $ larger than 1 nm \cite{Neuman2018}. As shown in Fig.~\ref{Figure1}(b), $\vec{\mu}=\left(\mu_{r},\mu_{\theta},\mu_{\phi} \right)$ is the dipole moment of the QE in the spherical coordinates. The strength of the dipole moment is $|\vec{\mu}| =0.38 $ e$\cdot$nm \cite{Delga2014}.

The Hamiltonian of the QE is given by $\hat{\emph{H}}_{\rm{QE}}=\hbar\left[\omega_{0}-i(\gamma_{\rm{QE}}/2)\right]\hat{\sigma}_{e_{1},e_{1}}$, where $\sigma _{e_{1},e_{1}}=|e_{1}\rangle \langle e_{1}|$, $|e_{1}\rangle$ is the excited state, and $\hbar \omega _{0}$ is the transition energy. Here, the Hamiltonian of the EM field can be expanded in terms of the annihilation (creation) operators of radiation field, $ \hat{f}(\vec{r},\omega)$ [$\hat{f}^{(\dagger)}(\vec{r},\omega)$], including all the EM modes of the vacuum and LSPs as $\hat{\emph{H}}_{\rm{EM}}=\int d^{3}\vec{r}\int_{0}^{\infty}d\omega\hbar\omega\hat{f}^{\dagger}(\vec{r},\omega)\hat{f}(\vec{r},\omega) $.
When excited, the QE is not only coupled electromagnetically to the LSPs on the metal surface, but also coupled to several decay channels, such as internal nonradiative decay due to rovibrational or phononic effects and spontaneous decay into the dielectric material, with a total rate $\gamma_{\rm{QE}}$. The interaction between the QE and EM modes is given by $\hat{\emph{H}}_{\rm{int}}=-\int_{0}^{\infty}
d\omega\{\vec{\mu}\cdot[\hat{E}(\vec{r},\omega)\hat{\sigma}^{(1)}_{+} +\rm{H.c.}]\}$, where $\hat{\sigma}_{+}^{(1)}=|e_{1}\rangle \langle g_{1}|$ represents the raising operator and $$\hat{E}(\vec{r},\omega)=i\sqrt{\frac{\hbar}{\pi\epsilon_{0}}}\frac{\omega^{2}}{c^{2}}\int d^{3}\vec{r}_{1}\sqrt{\epsilon^{\rm{I}}(\vec{r}_{1},\omega)}\widehat{G}(\vec{r},\vec{r}_{1},\omega)\hat{f}(\vec{r}_{1},\omega)$$ represents the quantized EM field \cite{Dung2000}. Here, $\epsilon^{\rm{I}}(\vec{r}_{1},\omega)$ is the imaginary part of $\epsilon(\vec{r}_{1},\omega)$, and $\rm{H.c.} $ stands for Hermitian conjugate. Note that $\widetilde{G}(\vec{r},\vec{r}_{1},\omega) $ represents the dyadic Green function obtained from the Maxwell-Helmholtz wave equation under the boundary condition
\begin{equation}
\begin{aligned}
\nabla\times\nabla\times\widetilde{G}(\vec{r},\vec{r}_{1},\omega)-\dfrac{\omega^{2}}{c^{2}}\epsilon(\vec{r},\omega)\widetilde{G}(\vec{r},\vec{r}_{1},\omega)=\textbf{I}\delta(\vec{r},\vec{r}_{1}), \label{eq0}
\end{aligned}
\end{equation}%
where $\textbf{I} $ stands for the unit dyad. In this regard, the Green's function, which contains all the information about the EM field in both dielectric and metal media, plays a prominent role in the realization of the coherent coupling between the QE and the MNP. Therefore, this QE-MNP system can then be described by the total Hamiltonian, within the rotating-wave approximation, as $\hat{\emph{H}}=\hat{\emph{H}}_{\rm{QE}}+\hat{\emph{H}}_{\rm{EM}}+\hat{\emph{H}}_{\rm{int}} $.
For a single quantum excitation, let $C_{1}(t)$ denote the probability amplitude that the QE can be excited. By solving the Schr\"odinger equation, one can obtain the following integro-differential equation for the $C_{1}(t)$ \cite{Tudela2014},
\begin{equation}
\begin{aligned}
\dfrac{d}{dt}C_{1}(t)=-\int_{0}^{t}dt_{1}\int_{0}^{\infty}d\omega \emph{J}(\omega)e^{i(\omega_{0}-\omega)(t-t_{1})}C_{1}(t_{1}), \label{eq1}
\end{aligned}
\end{equation}%
where $ \emph{J}(\omega) $ is the so-called spectral density of the QE-MNP system, which can be expanded into the sum of Lorentzian distributions (see Appendix.\ref{cal}),
\begin{equation}
\begin{aligned}
\emph{J}(\omega)&\approx\sum_{n=0}^{\infty}\dfrac{g_{n}^{2}}{\pi}\dfrac{\gamma_{\rm{p}}/2}{(\omega-\omega_{n})^{2}+(\gamma_{\rm{p}}/2)^{2}},\label{eq2}
\end{aligned}
\end{equation}%
where
\begin{equation}
\begin{aligned}
g_{n}^{2}=\sum_{\alpha=r,\theta,\phi}g_{n\alpha}^{2},\label{eq3}
\end{aligned}
\end{equation}%
with $$ \omega_{n}=\omega_{\rm{p}}/\sqrt{\epsilon_{\infty}+\epsilon_{\rm{d}}(n+1)/n}$$ being the cutoff frequency of the LSPs characterized by the angular momentum $ n $ and the Ohmic loss $ \gamma_{\rm{p}}$. Suppose that there is no direct tunneling between the MNP and the QE ($h>1 $ nm). Then the coupling strengths between the QE and the Lorentzian modes of the LSPs are given by
\begin{equation}
\begin{aligned}
g_{nr}^{2}=\mu_{r}^{2}(n+1)^{2}f_{n}(\omega_{n})
\end{aligned}
\end{equation}%
and
\begin{equation}
\begin{aligned}
g_{n\theta(\phi)}^{2}=\mu_{\theta(\phi)}^{2}\sum_{m=0}^{n}\mathcal{D}_{nm}\left[mP_{n}^{m}(0)\right]^{2}f_{n}(\omega_{n}),
\end{aligned}
\end{equation}%
where $$\mathcal{D}_{nm}=(2-\delta_{m0})\frac{(n-m)!}{(n+m)!},$$ $\delta_{m0}$ is the Kronecker delta function, $$f_{n}(\omega_{n})=\frac{a^{2n+1}}{(a+h)^{2n+4}}\left(1+\frac{1}{2n}\right)\frac{\omega_{\rm{p}}}{4\pi\epsilon_{0}\hbar}\left(\frac{\omega_{n}}{\omega_{\rm{p}}}\right)^{3}, $$ and $ P_{n}^{m}(x)$ is the associated Legendre polynomial. In order to properly evaluate and fit the polarization spectrum, the LSPs on the MNP can be approximately separated into the dipole mode and the pseudomode \cite{Hughes2018,Sebastian2018} with cutoff frequencies \cite{Delga2014} $ \omega_{\rm{d}}=\omega_{1} $ and $$ \omega_{\rm{M}}=\dfrac{\sum_{n=2}^{\infty}\omega_{n}g_{n}^{2}}{\sum_{n=2}^{\infty}g_{n}^{2}} $$ correspondingly. The couplings to the dipole mode and pseudomode are $g_{\rm{d}}=g_{1}$ and $g_{\rm{M}}^{2}=\sum_{n=2}^{\infty}g_{n}^{2} $, respectively.

Typically, the coupling to the pseudomode can be neglected when $h$ is large enough. However, when the QE is placed closer to the MNP with $h\leq 10$ nm, the coupling to the pseudomode can play a dominant role, even five to ten times stronger than the one to the dipole mode. In addition to $h$, both the coupling strengths, $g_{\rm{d}}$ and $g_{\rm{M}}$, also depend on the orientation of the transition dipole moment, as shown in Figs.~\ref{Figure1}(c) and \ref{Figure1}(d).

\begin{figure*}
\includegraphics[width=1.8\columnwidth]{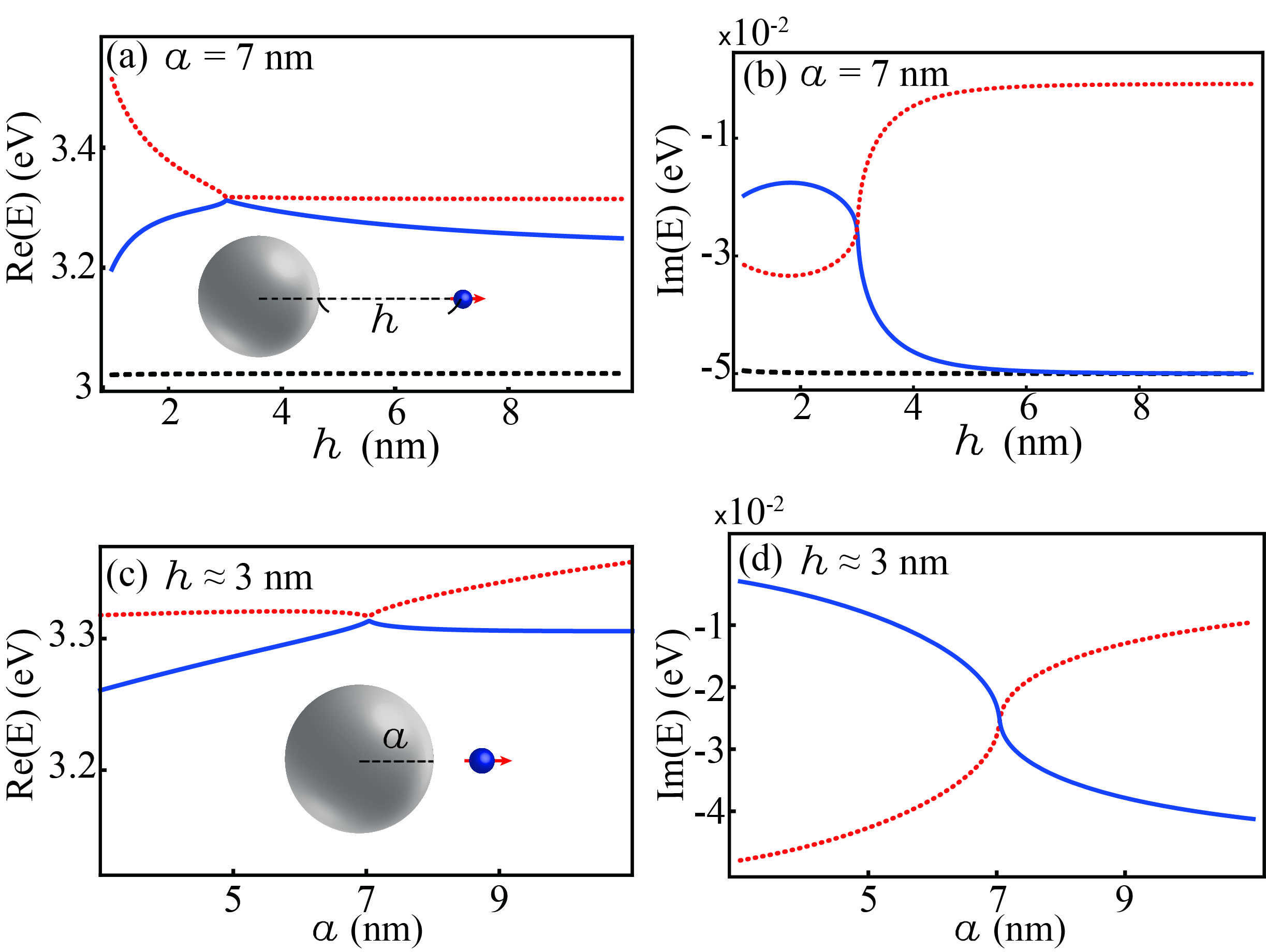}
%width=6.5cm,height=11cm%
\caption{Dependence of the eigenenergies of $H_{3\times 3}$ on the distance $h $ and the metal nanoparticle radius $a$. The dipole moment orientation is fixed at $(\mu,\theta_{\rm{d}},\phi_{\rm{d}})=(0.38,\pi/2,0)$ for the following calculations. (a) Real and (b) imaginary parts of the energy spectra jointly show the appearance of an exceptional point when placing the quantum emitter at $h\approx 3 $ nm. The occurrence of the eigenenergy splittings in the real (imaginary) part of the region, $h> 3$ ($h< 3$), is caused by both the detuning ($\omega_{\rm{M}}-\omega_{0}$) and the coupling to the dipole mode. (c) Real and (d) imaginary parts of energy spectra as a function of the metal nanoparticle radius $a $ showing an exceptional point.  }
\label{Figure2}
\end{figure*}

This QE-MNP system can formally be described by a non-Hermitian three-level Hamiltonian revealing EPs \cite{Delga2014},
\begin{equation}
\hat{H}_{3\times 3}=%
\begin{bmatrix}
\omega_{0}-i\frac{\gamma_{\rm{QE}}}{2} & g_{\rm{d}} & g_{\rm{M}}\\
g_{\rm{d}} & \omega_{\rm{d}}-i\frac{\gamma_{\rm{p}}}{2} & 0\\
g_{\rm{M}} & 0 & \omega_{\rm{M}}-i\frac{\gamma_{\rm{p}}}{2}%
\end{bmatrix}%
.\label{eq4}
\end{equation}%
When the QE is gradually moved closer to the MNP, both imaginary and real parts of the eigenenergies coalesce at a certain distance, resulting in the emergence of the EP. As shown in Figs.~\ref{Figure2}(a) and \ref{Figure2}(b), we can observe the EP while placing the QE at $h\approx 3$~nm with the dipole moment orientation $(\mu,\theta_{\rm{d}},\phi_{\rm{d}})=(0.38,\pi/2,0) $.

Due to the huge difference in the magnitudes, the appearance of an EP mainly results from the coupling between the QE and the pseudomode instead of the dipole mode. Consequently, to investigate the circumstance in which the EP forms, we consider a reduced Hamiltonian as well \citep{Choi2010,Kodigala2016}
\begin{equation}
\hat{H}_{2\times 2}=%
\begin{bmatrix}
\omega_{0}-i\frac{\gamma_{\rm{QE}}}{2} & g_{\rm{M}}\\
g_{\rm{M}} &  \omega_{\rm{M}}-i\frac{\gamma_{\rm{p}}}{2}%
\end{bmatrix}%
,
\end{equation}%
which describes the relevant coupling between the QE and the pseudomode; furthermore, it is a standard form of Hamiltonians generically studied in the context of EP \cite{Choi2010,Kodigala2016,Ozdemir2019}. It is clear that the EP can arise only under the conditions $\omega_{0}=\omega_{\rm{M}} $ and $g_{\rm{M}}=(\gamma_{\rm{p}}-\gamma_{\rm{QE}})/4 $.

In the vicinity of the EP shown in Fig.~\ref{Figure2}, the real parts of the eigenenergies drastically split when the QE is placed even closer to the MNP. It is noteworthy that, in the region where $h>3$, splitting occur as well, in contrast to the conventional coalesce observed in the previous literature \cite{Choi2010,Ganainy2018,Ozdemir2019}. These splittings are consequencies of the off-resonance condition $\omega_{0}\neq\omega_{\rm{M}}$ due to the dependence of $\omega_{\rm{M}}$ on $h$. Additionally, the coupling between the dipole mode and the QE also induces splitting. By solving the eigenvalue of $\hat{H}_{3\times 3}$ in Eq.~(\ref{eq4}), the splitting strength $\triangle E $, i.e., the difference between the dotted red and solid blue eigenenergies in Fig.~\ref{Figure2}, close to the EP can be analytically given by
\begin{equation}
\begin{aligned}
\triangle\mathrm{E}=\frac{-\sqrt{3}i\left[u^{2}+p(\omega_{\rm{d}},g_{\rm{d}})^{2/3}-48g_{\rm{d}}^{2} \right] }{12p(\omega_{\rm{d}},g_{\rm{d}})^{1/3}},\label{eq6}
\end{aligned}
\end{equation}%
where 
\begin{equation}
\begin{aligned}
p(\omega_{\rm{d}},g_{\rm{d}})=&144g_{\rm{d}}^{2}(2\omega_{\rm{d}\triangle}+i\gamma_{\triangle})+iu^{3}\\&+12q(\omega_{\rm{d}},g_{\rm{d}}),\label{p}
\end{aligned}
\end{equation}%
\begin{equation}
\begin{aligned}
q(\omega_{\rm{d}},g_{\rm{d}})=&-96g_{\rm{d}}^{4}(\gamma_{\triangle}^{2}-10i\gamma_{\triangle}\omega_{\rm{d}\triangle}+2\omega_{\rm{d}\triangle}^{2})
\\&-3\gamma_{\triangle}g_{\rm{d}}^{2}u^{3}-768g_{\rm{d}}^{6},\label{q}
\end{aligned}
\end{equation}%
$u=\gamma_{\triangle}+4i\omega_{\rm{d}\triangle}$, $\omega_{\rm{d}\triangle}=\omega_{\rm{d}}-\omega_{0}$, and $\gamma_{\triangle}=\gamma_{\rm{p}}-\gamma_{\rm{QE}}$. With Eq.~(\ref{eq6}), one can evaluate how the dipole mode coupling affects the eigenenergy splitting near an EP. Meanwhile, we find that the variations of the MNP radius can also achieve an EP, as shown in Figs.~\ref{Figure2}(c) and \ref{Figure2}(d).

\section{Detecting EP with power spectrum}\label{sec:model}
In engineering the presence of an EP in the QE-MNP system, one of the important issues is to first verify its existence. To this end, we propose to utilize the power spectrum as a potential means to do so \cite{Ridolfo2010} since it is experimentally measurable and usually exhibits features which are theoretically well understood%. and provides deep insights into the nature of light-matter interaction, particularly, in the field of quantum optics.

As we will see in the following, the behavior of the power spectrum reflects the features of the EP, including the coalesce of eigenenergies and the drastic splitting of corresponding eigenenergies near the EP. However, we will also show the difficulty of limited visibility due to the broadening in the power spectrum caused by dissipation.

By definition, the power spectrum $S(\omega)$ is given by
\begin{equation}
\begin{aligned}
S(\omega)=\frac{1}{\pi}\mathrm{Re}\int_{0}^{\infty}d\tau\langle\hat{\sigma}_{+}^{(1)}(0)\hat{\sigma}_{-}^{(1)}(\tau)\rangle e^{i \omega\tau},
\end{aligned}
\end{equation}%
where $\langle\hat{\sigma}_{+}^{(1)}(t)\hat{\sigma}_{-}^{(1)}(t+\tau)\rangle$ is the two-time correlation obtained by applying the quantum regression theorem to $\langle\hat{\sigma}_{-}^{(1)}(t)\rangle=\mathrm{Tr}[\hat{\sigma}_{-}^{(1)}\rho (t)]$. The time evolution of the QE-MNP system density matrix $\rho (t)$ is governed by the master equation
\begin{equation}
\begin{aligned}
\dot{\rho}(t)&=\dfrac{i}{\hbar}\left[\rho (t),\hat{\emph{H}}_{\rm{eff}}\right]+\dfrac{\gamma_{\rm{QE}}}{2}\mathcal{L}[\hat{\sigma}_{-}^{(1)}]\rho (t)\\&+\dfrac{\gamma_{\rm{p}}}{2}\sum_{\beta=\rm{d},\rm{M}}\mathcal{L}[\hat{a}_{\beta}]\rho (t)
\end{aligned}
\end{equation}%
with effective Hamiltonian
\begin{equation}
\begin{aligned}
\hat{\emph{H}}_{\rm{eff}}&=\hbar\omega_{0}\hat{\sigma}_{e_{1},e_{1}}\\&+\hbar\sum_{\beta=\rm{d},\rm{M}}\left[\omega_{\beta}\hat{a}_{\beta}^{\dagger}\hat{a}_{\beta}+g_{\beta}\left(\hat{a}_{\beta}\hat{\sigma}_{+}^{(1)}+\hat{a}_{\beta}^{\dagger}\hat{\sigma}_{-}^{(1)}\right) \right].
\end{aligned}
\end{equation}%
The superoperator $\mathcal{L}$ is defined as $$\mathcal{L}[\hat{o}]\rho (t)=\frac{1}{2}[2\hat{o}\rho (t)\hat{o}^{\dagger}-\rho (t)\hat{o}^{\dagger}\hat{o}-\hat{o}^{\dagger}\hat{o}\rho (t)].$$
\begin{figure*}
\centering
\includegraphics[width=2.05\columnwidth]{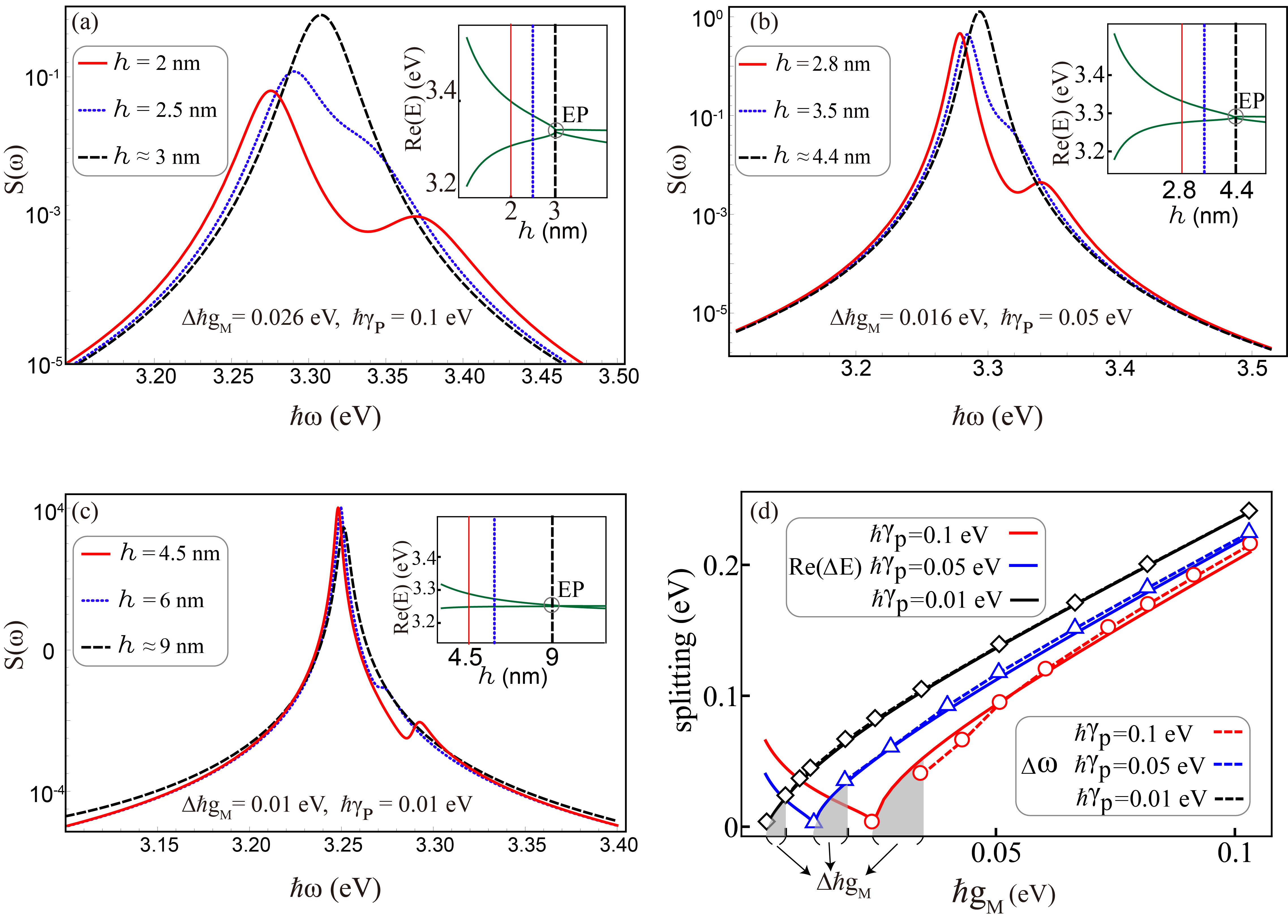}
%width=6.5cm,height=11cm%
\caption{(a) The power spectrum at different values of $h$ with $\hbar\gamma_{\rm{p}}=0.1$ eV. For the case of $h\approx 3$ nm (black dashed curve), we can observe a single main peak, which is consequence of the exceptional point, corresponding to the coalesce of eigenenergies shown in the inset. When moving the quantum emitter toward the metal nanoparticle at $h=2.5$ nm (blue dotted curve), it reaches the critical splitting and defines an increment threshold $\triangle\hbar g_{\rm{M}}=0.026$ eV, i.e., the difference of the $\hbar g_{\rm{M}}$ between $h\approx 3$ nm and $h=2.5$ nm. When the quantum emitter is even closer to the metal nanoparticle at $h=2$ nm (red solid curve), the splitting becomes more visible. In panels (b) and (c), in order to further investigate the relation between $\hbar\gamma_{\rm{p}}$ and  $\triangle\hbar g_{\rm{M}}$, the value of $\hbar\gamma_{\rm{p}}$ is reduced to $0.05$ eV and $0.01$ eV, respectively. The critical splittings occur at $h=3.5$ nm and $h=6$ nm, corresponding to the smaller thresholds $\triangle\hbar g_{\rm{M}}=0.016$ eV and $0.01$ eV, respectively. Additionally, it should be noted that, in these panels, the $\mathrm{S}(\omega)$-axis are shown on logarithmic scale. (d) The real part of the splitting strength $\rm{Re}(\triangle\rm{E})$ and the splitting in the power spectrum $\triangle\omega$ versus coupling strength $ \hbar g_{\rm{M}} $ with $\hbar\gamma_{\rm{p}}=0.1$ eV (red), 0.05 eV (blue), and 0.01 eV (black). The circle, triangle and square dots on the dashed curves represent the numerical data points for $\triangle\omega$. $\rm{Re}(\triangle\rm{E})$ rises drastically near the exceptional point. However, this behavior cannot be reflected in $\triangle\omega$ due to the broadening in the power spectrum. This leads to the undetectable regions marked in light gray; meanwhile, their widths equal to the threshold $\triangle\hbar g_{\rm{M}}$.}
\label{QE1gammapg1}
\end{figure*}

As shown by the black dashed curve in Fig.~\ref{QE1gammapg1}(a), corresponding to the presence of an EP at $ \hbar g_{\rm{M}}\approx 0.025$ eV, $h\approx 3$ nm, and $\hbar\gamma_{\rm{p}}=0.1$ eV, the single main peak is the consequence of the EP. When we gradually move the QE towards the MNP, a splitting is present in the energy spectrum, as shown by the green curves in the inset of Fig.~\ref{QE1gammapg1}(a). However, it should be noted that such splitting cannot be observed in the power spectrum until the QE is placed at $h=2.5$ nm (blue dotted curve). This obfuscation is due to the broadening caused by dissipation. As a result of the dependence on $h$, the coupling strength $g_{\rm{M}}$ increases during the QE movement, and, consequently, we can define an increment threshold $ \triangle\hbar g_{\rm{M}}=0.026$ eV, i.e., the difference of coupling strength between the emergence of an EP and the beginning of the splitting. As the QE is positioned at $h=2$ nm (red solid curve), the splitting is even more notable since the increment in $ \hbar g_{\rm{M}}$ significantly exceeds the threshold $ \triangle\hbar g_{\rm{M}}$.

This implies that the visibility of the splitting in the power spectrum is an intuitive benchmark of its performance in detecting the EP. To enhance the visibility, it is critical to suppress the broadening in the power spectrum, such that the increment in $\hbar g_{\rm{M}}$ can more easily exceed the threshold $ \triangle\hbar g_{\rm{M}}$. Doing so helps us to rule out the region where the EP has been broken and, in turn, to pin down a smaller parameter range containing the EP, and hence improve the sensitivity.

As the dissipation is the origin of the broadening in the power spectrum, the former is responsible for the value of the threshold $ \triangle\hbar g_{\rm{M}}$ as well. In order to investigate the relation between $\hbar\gamma_{\rm{p}}$ and $ \triangle\hbar g_{\rm{M}}$, we further reduce the value of $\hbar\gamma_{\rm{p}}$ to $0.05$ eV and $0.01$ eV in Figs.~\ref{QE1gammapg1}(b) and \ref{QE1gammapg1}(c), respectively. Following the same analysis as Fig.~\ref{QE1gammapg1}(a), we conclude that the corresponding $ \triangle\hbar g_{\rm{M}}=0.016$ eV and $0.01$ eV, respectively, in line with our explanation above.

To further schematically elaborate the intimate relation between $\hbar\gamma_{\rm{p}}$ and $ \triangle\hbar g_{\rm{M}}$, in Fig.~\ref{QE1gammapg1}(d), we depict the real part of the splitting strength, $\rm{Re}(\triangle E)$ (solid curves), of Eq.~(\ref{eq6}) and the visible splitting, $\triangle\omega$ (dashed curve), in the power spectrum, at $\hbar\gamma_{\rm{p}}=0.1$ eV, $0.05$ eV, and $0.01$ eV. Although the splitting in the energy spectrum around the EP is drastic, it cannot be reflected by $\triangle\omega$, due to the dissipation-induced broadening, as explained above. $\triangle\omega$ is finite only if the increment in $\hbar g_{\rm{M}}$ exceeds $\triangle\hbar g_{\rm{M}}$. This leads to the undetectable region, marked by the light gray areas in Fig.~\ref{QE1gammapg1}(d). It is clear that the smaller the $\gamma_{\rm{p}}$, the smaller the increment threshold $ \triangle\hbar g_{\rm{M}}$.

\section{Exceptional points induced by collective coupling to surface plasmons}
\begin{figure}
\includegraphics[width=0.48\textwidth]{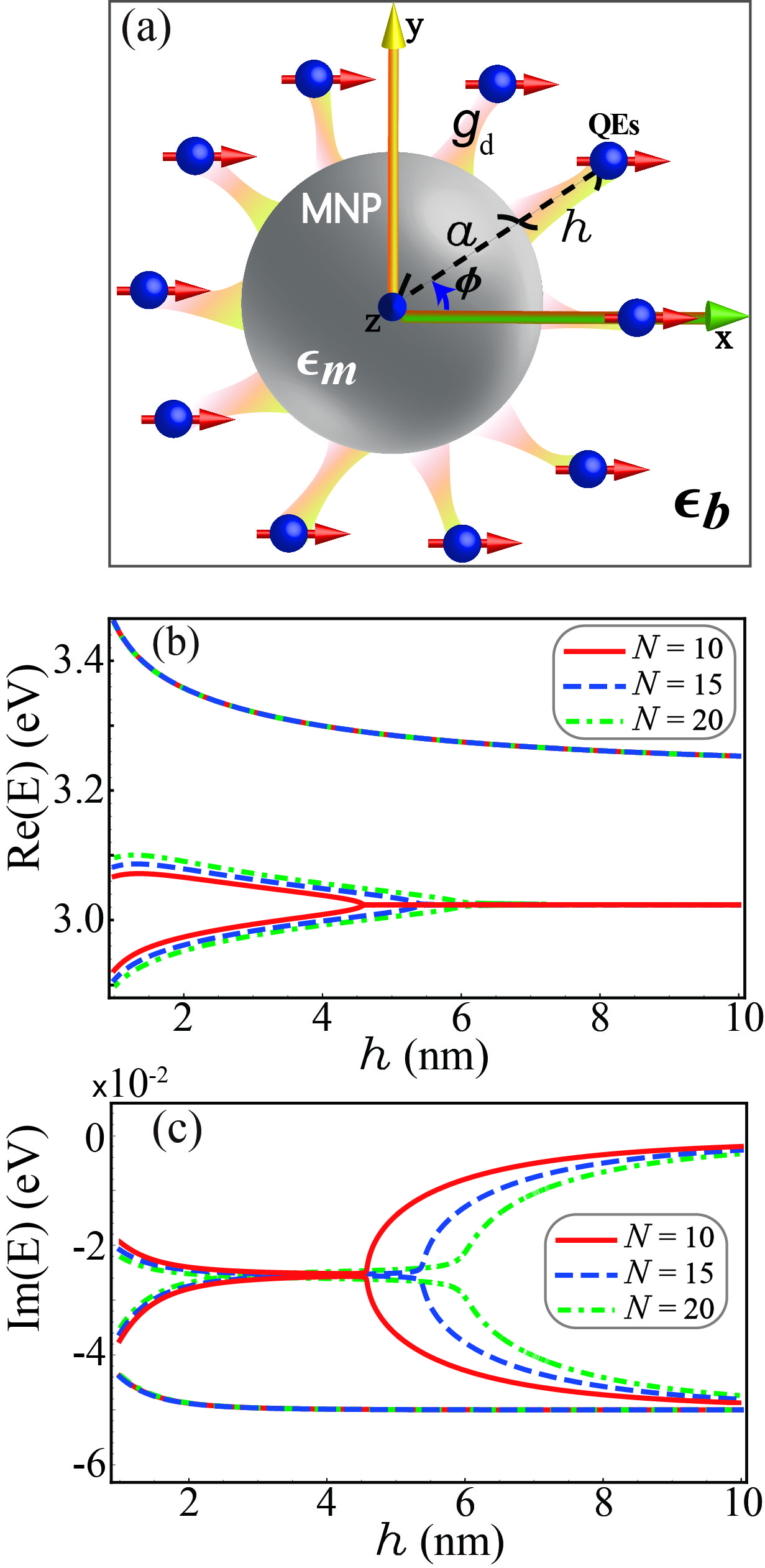}
%width=2.166cm,height=3.6cm%
\caption{(a) Schematic of $N$ quantum emitters placed near the metal nanoparticle on the $x$-$y$ plane at an identical distance $h$ from the metal nanoparticle surface. Each quantum emitter is separated from one another by $\phi=2\pi/N$ along the $\vec{\phi}$ direction, with the dipole moment orientation being parallel to $x$ axis. The real (b) and imaginary (c) parts of the energy spectra as a function of $h$ with $N=10$ (red), $N=15$ (blue) and $N=20$ (green) quantum emitters. The observable shift of the exceptional point position, as well as the eigenenergy splitting, emerge when increasing the number of quantum emitters. The noticeable eigenenergy splittings result from an immense strength of the dipole-dipole interaction due to the close distances among quantum emitters. Note that by increasing the number of quantum emitters, a stronger collective coupling to the dipole mode can be observed, such that the exceptional point occurs when the quantum emitters are placed at a longer distance $h_{\rm{EP}}$ with respect to the surface of the metal nanoparticle.}
\label{Figure3}
\end{figure}

When increasing the number of QEs near the MNP, one can expect a stronger interaction between the dipole mode of LSPs and the QEs compared to the previous case \cite{DelgaVidal2014}. Hence, the strong-coupling regime between the LSPs and QEs can be easily reached by their collective coupling to the dipole mode, rather than to the pseudomode. In this regard, an EP is likely to be achieved via the collective coupling between the dipole mode and the QEs. We assume that there are $N$ QEs arranged radially at $h$ nm from the surface of the MNP with an identical dipole moment orientation of each $i$'s QE $\vec{\mu}_{i}=(\mu_{r},\mu_{\theta},\mu_{\phi})$ being parallel to $x$ axis, as illustrated in Fig.~\ref{Figure3}(a).

Transforming the dipole-dipole interaction $$J_{ij}=\frac{1}{4\pi\epsilon_{b}}[\vec{\mu}_{i}\cdot\vec{\mu}_{j}/\vert\vec{r}_{ij}\vert^{3}-3(\vec{\mu}_{i}\cdot\vec{r}_{ij})(\vec{\mu}_{j}\cdot\vec{r}_{ij})/\vert\vec{r}_{ij}\vert^{5}] $$ into the effective detuning $\delta_{J}$ with identical distance $r_{ij}$ between each adjacent QEs, the interactions between the QEs and the LSPs can be described by the three-level non-Hermitian Hamiltonian
\begin{equation}
H_{3\times 3}=%
\begin{bmatrix}
\omega_{0}-i\frac{\gamma_{\rm{QE}}}{2}+\delta_{\rm{J}} & \sqrt{N}g_{\rm{d}} & g_{\rm{M}}\\
\sqrt{N}g_{\rm{d}} & \omega_{\rm{d}}-i\frac{\gamma_{\rm{p}}}{2} & 0\\
g_{\rm{M}} & 0 & \omega_{\rm{M}}-i\frac{\gamma_{\rm{p}}}{2}%
\end{bmatrix}%
.
\end{equation}%
Note that a similar Hamiltonian was studied in Refs.~\cite{DelgaVidal2014,Delga2014} but not in the context of EPs. From the eigenvalue equations, one can obtain the energy spectra with the emergence of an EP by setting ten to twenty QEs in proximity to MNP as depicted in Figs.~\ref{Figure3}(b) and \ref{Figure3}(c).
\begin{figure}[htbp]
\includegraphics[width=0.47\textwidth]{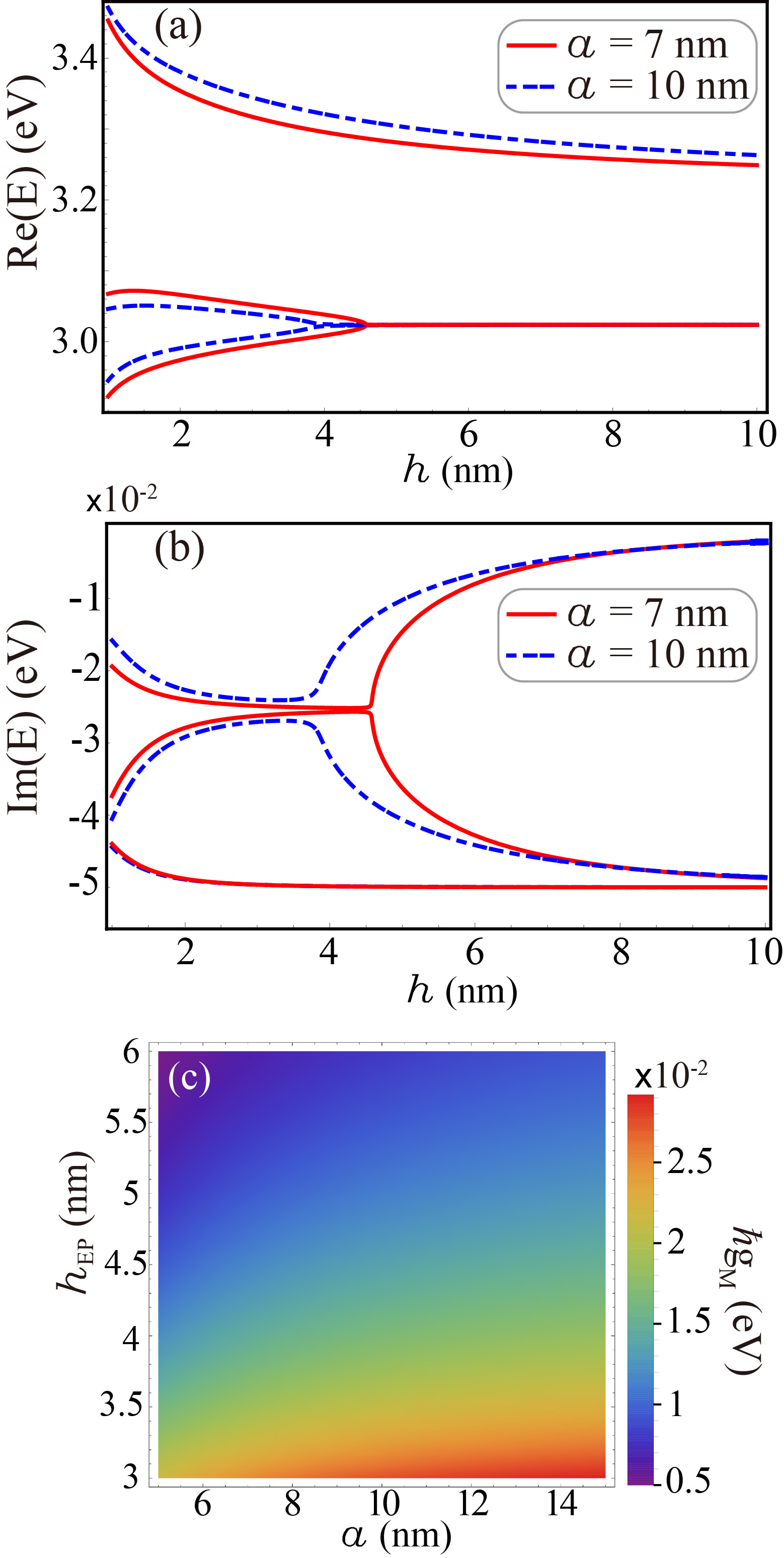}
%width=2.166cm,height=3.6cm%
\caption{Real (a) and imaginary (b) parts of the energy spectra as a function of distance $h$ with the radius $a=7$ (red solid) and $a=10$ (blue dashed) when setting the number of quantum emitters $N=10$. The larger the radius $a$, the smaller the distance $h_{\rm{EP}}$, where an exceptional point emerges. (c) The eigenenergy splittings are triggered by an enhanced coupling to the pseudomode $g_{\rm{M}} $ with both increment of $a$ and reduction in $h_{\rm{EP}}$, which indicates that the region in blue color is suitable to perform the exceptional point. }
\label{Figure4}
\end{figure}
Therefore, compared to the single-QE case, an EP here is mainly triggered by the collective coupling between the QEs and the dipole mode instead of the pseudomode. As we increase the number of the QEs, an EP occurs when the QEs are placed at a further distance $h_{\rm{EP}}$ from the surface of the MNP. However, if the composite system contains 20 QEs or even more, a complete energy splitting can occur due to the strong dipole-dipole interaction, such that the EP disappears. Therefore, there is a limit on the suitable number of the QEs to achieve an EP in the energy spectrum.

Besides the dipole-dipole interaction, the QEs coupled to the pseudomode also play an important role in the formation of eigenenergy splitting near an EP. In order to investigate how the coupling to the pseudomode affects the eigenenergy splitting, we make a comparison between different sizes of the MNPs coupled to 10 QEs with the significant enhancement of the coupling to the pseudomode. The splitting emerges while enlarging the MNP size from $a=7 $ to $10$ nm, as shown in Figs.~\ref{Figure4}(a) and \ref{Figure4}(b). This is because the enlargement of the MNP size is beneficial to enhance the coupling to the pseudomode.

Meanwhile, the distance $h_{\rm{EP}}$ for the occurrence of EP becomes closer to the MNP. In this case, the coupling to the pseudomode can be enhanced, thereby increasing the splitting simultaneously. As shown in Figs.~\ref{Figure4}(c) and \ref{Figure4}(d), the relation between $h_{\rm{EP}}$ and the radius $a $ of the MNP, can be described by an analytic form
\begin{equation}
h_{\rm{EP}}=-a+\left[\frac{6N\omega_{\rm{d}}^{3}(4\mu_{r}^{2}+\mu_{\theta}^{2}+\mu_{\phi}^{2})}{\pi\epsilon_{0}\omega_{\rm{p}}^{2}(\gamma_{\rm{p}}-\gamma_{\rm{QE}})^{2}}\right]^{\frac{1}{6}}\sqrt{a}.
\end{equation}%
Overall, the enlargement of the eigenenergies splitting at the position of an EP results from the enhancement of the coupling to the pseudomode, which can be caused by both increase of the MNP size and reduction in the distance $h_{\rm{EP}}$.

\section{Detecting the exceptional points of numerous quantum emitters case}
\begin{figure*}
\includegraphics[width=2.1\columnwidth]{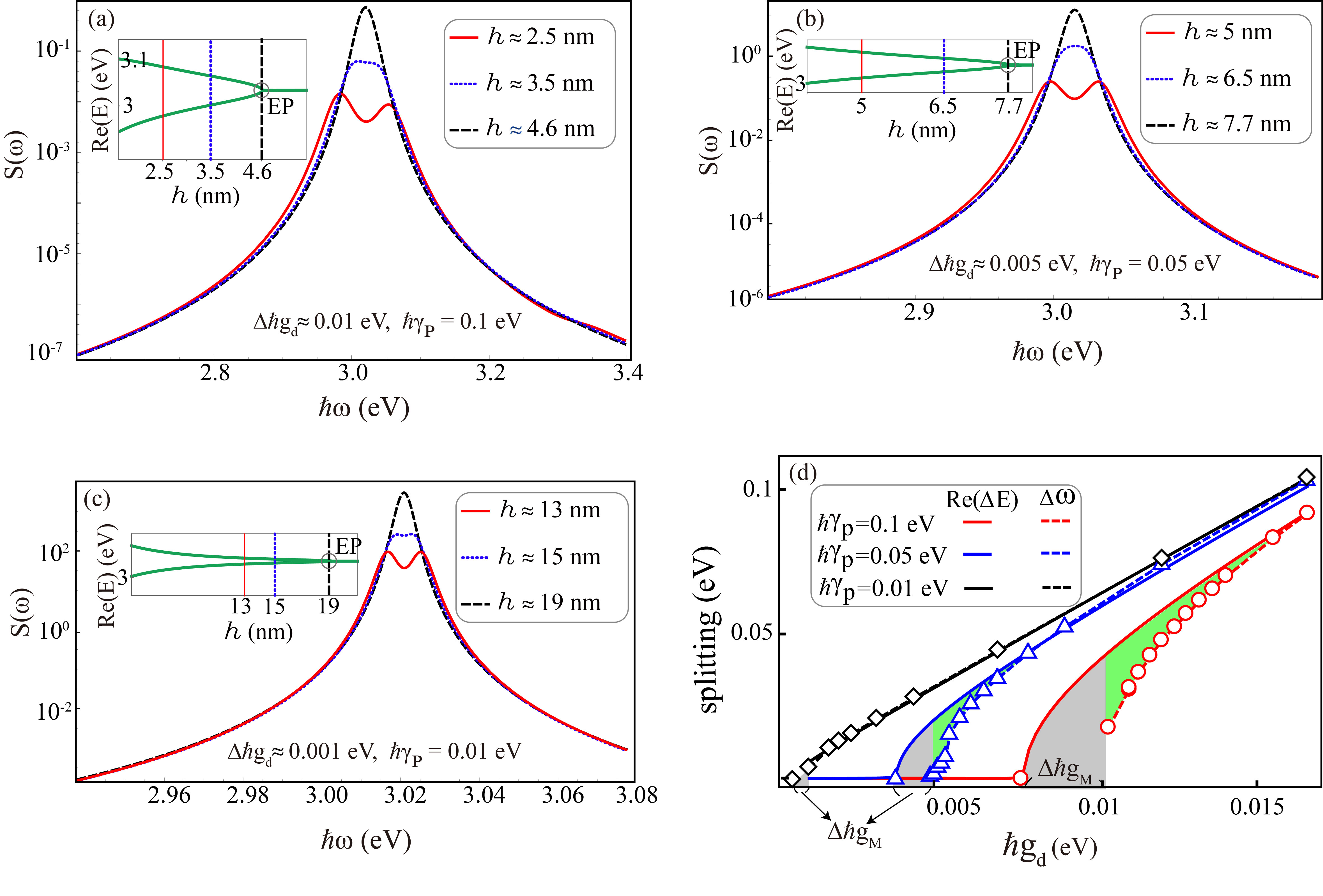}
%width=6.5cm,height=11cm%
%\centering
\caption{(a) The power spectrum of 10 quantum emitters for different values of $h$ with $\hbar\gamma_{\rm{p}}=0.1$ eV. For the case of $h\approx 4.6$ nm (black dashed curve), the observation of a single main peak is a consequence of the exceptional point, corresponding to the coalesce of eigenenergies shown in the inset. When moving the quantum emitter toward metal nanoparticle at $h=3.5$ nm (blue dotted curve), a splitting starts to emerge from the broadening, which is the critical splitting, and here is defined as an increment threshold $\triangle\hbar g_{\rm{d}}\approx0.01$ eV. Moreover, when the quantum emitter is even closer to the metal nanoparticle at $h=2.5$ nm (red solid curve), the splitting becomes more noticeable. In panels (b) and (c), in order to probe the relation between $\hbar\gamma_{\rm{p}}$ and  $\triangle\hbar g_{\rm{d}}$, the value of $\hbar\gamma_{\rm{p}}$ is reduced to $0.05$ eV and $0.01$ eV, respectively. The critical splittings occur at $h=6.5$ nm and $h=15$ nm, with respect to the smaller threshold $\triangle\hbar g_{\rm{d}}=0.005$ eV and $0.001$ eV, correspondingly. Additionally, it should be noted that, in these panels, the $\mathrm{S}(\omega)$-axis are shown on logarithmic scale. (d) The real part of the splitting strength $\rm{Re}(\triangle\rm{E})$ and the splitting in the power spectrum $\triangle\omega$ versus coupling strength $ \hbar g_{\rm{d}} $ with $\hbar\gamma_{\rm{p}}=0.1$ eV (red), 0.05 eV (blue), and 0.01 eV (black). The circle, triangle and square dots on the dashed curves represent the numerical data points for $\triangle\omega$. Although the $\triangle\omega$ near exceptional point in the case of $\hbar\gamma_{\rm{p}}=0.1$ eV possess a stronger dependence on $\hbar g_{\rm{d}} $ than the case of $\hbar\gamma_{\rm{p}}=0.05$ eV, however, it comes with the larger threshold $\triangle\hbar g_{\rm{d}}$ (gray areas) in proximity to the exceptional point due to the wider broadening in the power spectrum. In addition, the green areas show that the $\triangle\omega$ in the power spectrum are smaller than the expected strength in the energy spectrum due to the larger $\hbar\gamma_{\rm{p}}$.
}
\label{10QEgammapgvsomega1}
\end{figure*}
Once again we can utilize the power spectrum to detect the presence of the EP in a system composed of numerous QEs coupled to LSP.  To do so we first consider the master equation of QEs-MNP system
\begin{equation}
\begin{aligned}
\dot{\rho} (t)&=\dfrac{i}{\hbar}\left[\rho (t),\hat{\emph{H}}_{\rm{eff}}\right]+\dfrac{\gamma_{\rm{QE}}}{2}\mathcal{L}[\hat{\sigma}_{-}^{(c)}]\rho (t)\\&+\dfrac{\gamma_{\rm{p}}}{2}\sum_{\beta=\rm{d},\rm{M}}\mathcal{L}[\hat{a}_{\beta}]\rho (t),
\end{aligned}
\end{equation}%
with effective Hamiltonian
\begin{equation}
\begin{aligned}
\hat{\emph{H}}_{\rm{eff}}&=\hbar(\omega_{0}+\delta_{\rm{J}})\hat{\sigma}_{+}^{(c)}\hat{\sigma}_{-}^{(c)}+\hbar\sum_{\beta=\rm{d},\rm{M}}\omega_{\beta}\hat{a}_{\beta}^{\dagger}\hat{a}_{\beta}\\&+\hbar\sqrt{N}g_{\rm{d}}\left(\hat{a}_{\rm{d}}\hat{\sigma}_{+}^{(c)}+\hat{a}_{\rm{d}}^{\dagger}\hat{\sigma}_{-}^{(c)}\right)\\&+\hbar g_{\rm{M}}\left(\hat{a}_{\rm{M}}\hat{\sigma}_{+}^{(c)}+\hat{a}_{\rm{M}}^{\dagger}\hat{\sigma}_{-}^{(c)}\right),
\end{aligned}
\end{equation}%
where the $\hat{\sigma}_{+}^{(c)}(\hat{\sigma}_{-}^{(c)})$ represents the raising (lowering) operator for collection of QEs. Following the same procedure used in the single QE case, by calculating two-time correlations $\langle\hat{\sigma}_{+}^{(c)}(t)\hat{\sigma}_{-}^{(c)}(t+\tau)\rangle$, one can obtain the power spectrum $$S(\omega)=\frac{1}{\pi}\mathrm{Re}\int_{0}^{\infty}d\tau\langle\hat{\sigma}_{+}^{(c)}(0)\hat{\sigma}_{-}^{(c)}(\tau)\rangle e^{i \omega\tau}.$$

Under suitable conditions, an EP is exhibited, i.e., $ \sqrt{10}\hbar g_{\rm{d}}\approx 0.025$ eV, $h\approx 4.6$ nm, $\hbar\gamma_{\rm{p}}=0.1$ eV, and we can observe only one main peak in the power spectrum as shown by the black dashed curve in Fig.~\ref{10QEgammapgvsomega1}(a), corresponding to the coalesce of eigenenergies, as shown by the green curves in the inset. When we move the QE toward the MNP, a drastic splitting is present in the energy spectrum. However, such splitting cannot be observed readily in the power spectrum since another peak merges into the main peak. Using the same approach as in the single QE case, the QEs should be placed at a close enough distance, i.e., $h=3.5$ nm (blue dotted curve), in order to exceed the threshold of coupling to the dipole mode $ \triangle\hbar g_{\rm{d}}=0.01$ eV. As the QE is positioned at $h=2$ nm (red solid curve), the splitting is more notable since the increment in $ \hbar g_{\rm{d}}$ significantly exceeds the threshold $ \triangle\hbar g_{\rm{d}}$.

Analogous to the single QE case, in order to explore the relation between $\hbar\gamma_{\rm{p}}$ and $ \triangle\hbar g_{\rm{d}}$ for 10 QEs case, we then reduce the value of $\hbar\gamma_{\rm{p}}$ to $0.05$ eV and $0.01$ eV in Figs.~\ref{10QEgammapgvsomega1}(b) and \ref{10QEgammapgvsomega1}(c), corresponding to a threshold $ \triangle\hbar g_{\rm{d}}=0.005$ eV and 0.001 eV, respectively. As expected, the QEs-MNP system with a smaller dissipation requires the smaller $ \triangle\hbar g_{\rm{d}}$ to observe the splitting.

Therefore, to further schematically analyze the relation between $\hbar\gamma_{\rm{p}}$ and $ \triangle\hbar g_{\rm{d}}$, we plot the real part of the eigenenergies splitting near an EP, $\rm{Re}(\triangle E)$ (solid curves), and also the visible splitting, $\triangle\omega$ (dashed curve) in the power spectrum,  at $\hbar\gamma_{\rm{p}}=0.1$ eV, $0.05$ eV, and $0.01$ eV, in Figs.~\ref{10QEgammapgvsomega1}(d). It is easier to identify the presence of EP via the variation of $g_{\rm{d}}$ if the splitting near an EP is strongly dependent on $g_{\rm{d}}$. This is because, under the influence of larger dissipation, the QE-MNP system possesses more drastic splitting near the EP with respect to $g_{\rm{d}}$ in the energy spectrum. However, for the larger dissipation case in the power spectrum, rather than the clearer observation of the drastic splitting, it actually comes with not only the larger threshold $\triangle\hbar g_{\rm{d}}$ (gray area), but also a smaller splitting than the expected value in the energy spectrum (green area) due to the larger $\hbar\gamma_{\rm{p}}$.

\begin{figure}
\includegraphics[width=1\columnwidth]{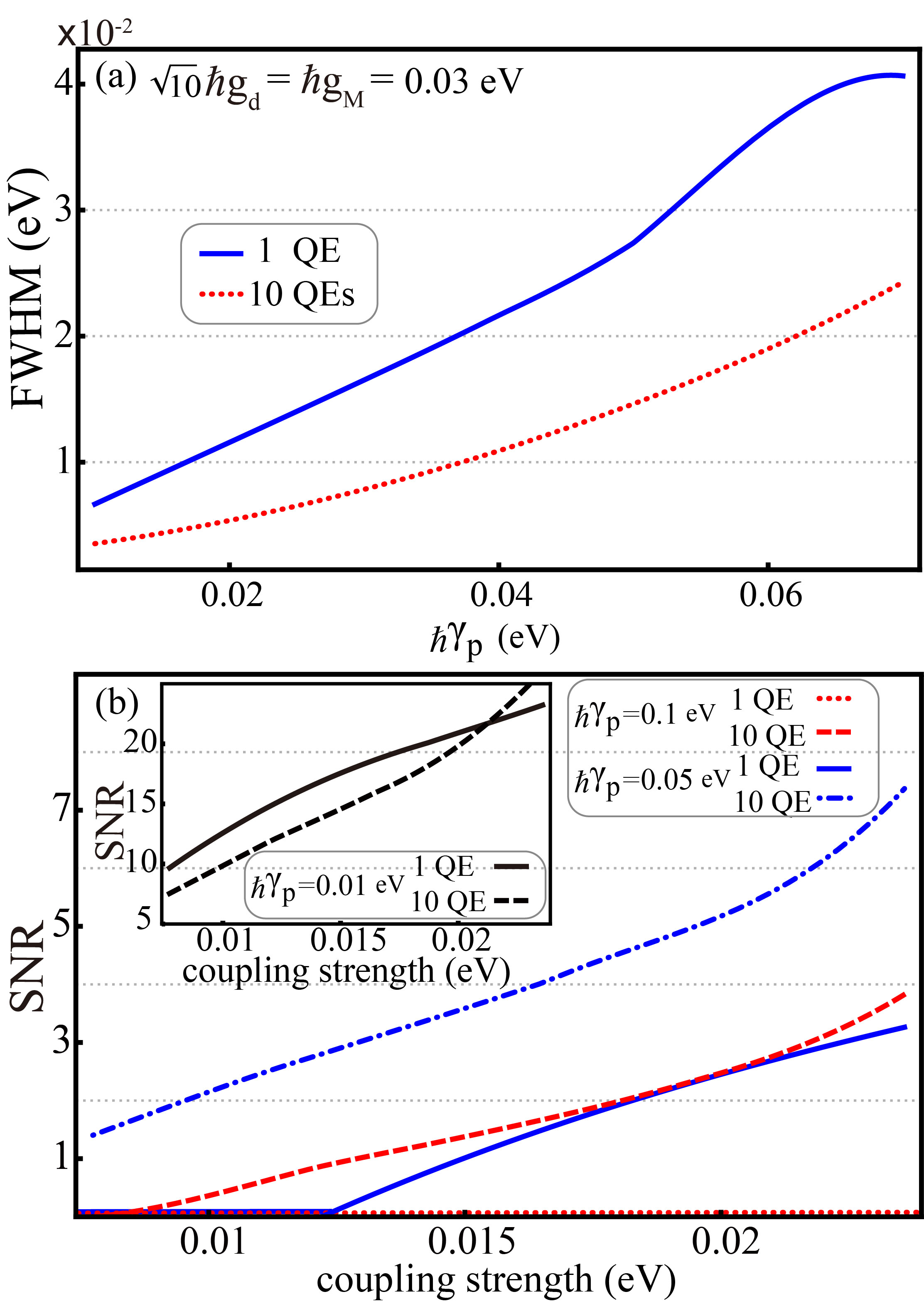}
%width=6.5cm,height=11cm%
\caption{(a) Variation in full width at half maximum (FWHM) versus dissipation $\hbar\gamma_{\rm{p}}$. Solid blue and dotted red curves stand for the cases of single quantum emitter and 10 quantum emitters, respectively. The FWHM has positive correlation with $\hbar\gamma_{\rm{p}}$. (b) signal-to-noise ratio (SNR) versus coupling strength of a individual quantum emitter. For $\hbar\gamma_{\rm{p}}=0.01$ eV case shown in the inset, the SNR of single quantum emitter case (solid black curve) is slightly higher than 10 quantum emitters case (dashed black curve) within the range of $\hbar g_{\rm{d}(\rm{M})}=0$ to 0.02 eV. On the contrary, for the larger dissipation $\hbar\gamma_{\rm{p}}=0.05$ eV (solid and dot-dashed blue curves stand for the cases of single quantum emitter and 10 quantum emitters, respectively.) and $\hbar\gamma_{\rm{p}}=0.1$ eV (dotted and dashed red curves stand for the cases of single quantum emitter and 10 quantum emitters, respectively.), the SNRs of 10 quantum emitters case are both higher than single quantum emitter case.}
\label{SNR1}
\end{figure}

For this reason, how accurately the EP can be observed depends on how large of a strength of splitting one can detect near the EP in power spectrum. Thus, the splitting ($\triangle\omega$) here is the information to be regarded as `the signal'. The width of the main peak will obscure the ability to extract the required splitting. For convenience, the width of the main peak can be quantified using the full width at half maximum (FWHM), which has a relationship with $\hbar\gamma_{\rm{p}}$ as depicted in Fig.~\ref{SNR1}(a). In this regard, we thus define a signal-to-noise ratio (SNR), accounting for the splitting and FWHM, which quantifies the resolution of power spectrum
\begin{equation}
\begin{aligned}
\mathrm{SNR}=\dfrac{\triangle\omega}{\mathrm{FWHM}}.
\end{aligned}
\end{equation}%
One can infer that the higher the SNR, the better the resolution for detecting the EPs. Comparing the SNR of the single QE case to the 10 QEs case, in terms of coupling strength, we find  that although the SNR of a single QE case is slightly higher than 10 QEs case for $\hbar\gamma_{\rm{p}}=0.01$ eV, the SNR of 10 QEs case is conversely greater for both $\hbar\gamma_{\rm{p}}=0.05$ eV and $0.1$ eV as shown in Fig.~\ref{SNR1}(b).
Combining these results, a collection of QEs possesses better SNR for larger system dissipation when considering the individual coupling strength of each QE.
%--------------------------------------------------------------------------------------------------------------------------------------

\section{CONCLUSIONS}\label{sec:conclusion}
In conclusion, we have shown the emergence of EPs in open quantum systems composed of an MNP and a number of QEs. Surprisingly, an EP can stem from the coupling of different modes between a QE and LSPs. For the single-QE case, the formation of an EP mainly results from the coupling to a pseudomode, which becomes dominant when the distance between the QE and the MNP is shorter than 10 nm. However, the coupling to the dipole mode plays an important role in inducing the eigenenergy splittings near an EP. Subsequently, placing more QEs nearby the MNP triggers a collective coupling, which also induces an EP. Instead of the coupling to the dipole mode, the coupling to the pseudomode and the dipole-dipole interaction between QEs become important factors leading the splitting of eigenenergies near an EP. Therefore, with a proper balance between the quantum collective effect and the dipole-dipole interaction, by using a number of QEs near the MNP not only relaxes the strong-coupling requirement for an individual QE, but also results in a more stable condition to generate exceptional points.

Furthermore, we have shown that EPs can be revealed in power spectra. Specifically, EPs correspond to frequency splitting in a power spectrum. We find that the system's dissipation sets a detection limit of observable splitting near an EP. The SNR analysis, accounting for frequency splitting and the system's dissipation, enables us to evaluate the accuracy of the observation of EPs. We conclude that a collection of QEs coupled to an MNP offers unique advantages in terms of a better SNR for larger system's dissipation compared to that of a single QE.

%-------------------------------------------------------------------------------------------------------------------------------------
\section*{ACKNOWLEDGMENTS}\label{sec:acknowledgments}
Y.N.C. acknowledges the support of the Ministry of Science and Technology, Taiwan (Grants No. MOST 107-2628-M-006-002-MY3 and MOST 107-2627-E-006-001), and Army Research Office (Grant No. W911NF-19-1-0081). H.B.C. acknowledges the support of the Ministry of Science and Technology, Taiwan (Grant No. MOST 108-2112-M-006-020-MY2). G.Y.C. acknowledges the support of the Ministry of Science and Technology, Taiwan (Grant No. MOST 105-2112-M-005-008-MY3). F.N. is supported in part by the: MURI Center for Dynamic Magneto-Optics via the Air Force Office of Scientific Research (AFOSR) (FA9550-14-1-0040), Army Research Office (ARO) (Grant No. Grant No. W911NF-18-1-0358), Asian Office of Aerospace Research and Development (AOARD) (Grant No. FA2386-18-1-4045), Japan Science and Technology Agency (JST) (via the Q-LEAP program, and the CREST Grant No. JPMJCR1676), Japan Society for the Promotion of Science (JSPS) (JSPS-RFBR Grant No. 17-52-50023, and JSPS-FWO Grant No. VS.059.18N), the Foundational Questions Institute (FQXi), and the NTT PHI Laboratory.  N.L. and F.N. acknowledge support from the RIKEN-AIST Challenge Research Fund.  N.L., A.M., Y.N.C., and F.N. are supported by the Sir John Templeton Foundation.  N.L. acknowledges additional support from JST PRESTO, Grant No. JPMJPR18GC.

%---------------------------------------------------------------------------------------------------------

\onecolumngrid
\appendix
\section{Derivations of the spectral density in the quasistatic limit}\label{cal}

From the integro-differential equation, given by Eq.~(\ref{eq1}), one can obtain the spectral density as follows \cite{Tudela2014}
\begin{equation}
\begin{aligned}
\emph{J}(\omega)=\frac{\omega^{2}}{\pi\epsilon_{0}c^{2}}\vec{\mu}_{1}\cdot\mathrm{Im}[\widehat{G}(\vec{r}_{1},\vec{r}_{1},\omega)]\cdot\vec{\mu}_{1},\label{eqs1}
\end{aligned}\tag{A1}
\end{equation}%
where the Green's tensor $\widehat{G}(\vec{r}_{1},\vec{r}_{1},\omega)$ satisfy the boundary conditions \cite{Li1994}
\begin{equation}
\begin{aligned}
\widehat{G}(\vec{r}_{1},\vec{r}_{1},\omega)=\widehat{G}_{0}(\vec{r}_{1},\vec{r}_{1},\omega)+\widehat{G}_{\rm{scatt}}(\vec{r}_{1},\vec{r}_{1},\omega).
\end{aligned}\tag{A2}
\end{equation}%
Here, the full Green's tensor is composed of the unbounded dyadic Green's function, $\widehat{G}_{0}(\vec{r}_{1},\vec{r}_{1},\omega)$ and the scattering dyadic Green's function, $\widehat{G}_{\rm{scatt}}(\vec{r}_{1},\vec{r}_{1},\omega)$, which represent the vacuum contribution and an additional contribution of the multiple reflection and transmission waves, respectively. Given a silver MNP with the wave vector $k_{m}=\frac{\omega}{c}\sqrt{\epsilon_{m}}$ embedded in a homogeneous medium of wave vector $k_{b}=\frac{\omega}{c}\sqrt{\epsilon_{b}}$, the unbounded part of the dyadic Green's function, in terms of the spherical vector wave functions, is given by
\begin{equation}
\begin{aligned}
\widehat{G}_{0}(\vec{r}_{1},\vec{r}_{1},\omega)=k_{b}\sum_{s=\pm}\sum_{n=0}^{\infty}\sum_{m=0}^{n}f_{mn}[\vec{M}_{mn}^{s(1)}(k_{b}r_{1})\vec{M}_{mn}^{s}(k_{b}r_{1})+\vec{N}_{mn}^{s(1)}(k_{b}r_{1})\vec{N}_{mn}^{s}(k_{b}r_{1})],
\end{aligned}\tag{A3}
\end{equation}%
and the scattered part of the dyadic Green function can be expanded as
\begin{equation}
\begin{aligned}
\widehat{G}_{\rm{scatt}}(\vec{r}_{1},\vec{r}_{1},\omega)=k_{b}\sum_{s=\pm}\sum_{n=0}^{\infty}\sum_{m=0}^{n}f_{mn}[R_{H}\vec{M}_{mn}^{s(1)}(k_{b}r_{1})\vec{M}_{mn}^{s(1)}(k_{b}r_{1})+R_{V}\vec{N}_{mn}^{s(1)}(k_{b}r_{1})\vec{N}_{mn}^{s(1)}(k_{b}r_{1})],
\end{aligned}\tag{A4}
\end{equation}%
where $$f_{mn}=\frac{i}{4\pi}\left(2-\delta_{m0}\right)\frac{2n+1}{n(n+1)}\frac{(n-m)!}{(n+m)!}.$$ Here, $R_{H}$ and $R_{V}$ represent the centrifugal reflection coefficients corresponding to the electric field of the TE and TM waves, respectively. We take the quasistatic limit into account due to the justified assumption that the distance between the MNP and QE is much smaller than the wavelength of the  electromagnetic field $(k_{b}r_{1}\ll 1)$. Hence, the values of $R_{H}$ and $R_{V}$ are given by
\begin{equation}
\begin{aligned}
R_{H}\approx\frac{i\pi a^{2n+3}k_{b}^{2n+1}(k_{m}^{2}-k_{b}^{2})}{2^{2n+4}\Gamma(n+\frac{3}{2})\Gamma(n+\frac{5}{2})},
\end{aligned}\tag{A5a}
\end{equation}%
\begin{equation}
\begin{aligned}
R_{V}\approx\frac{i\pi (n+1)(k_{b}a)^{2n+1}(k_{m}^{2}-k_{b}^{2})}{2^{2n+1}\Gamma(n+\frac{1}{2})\Gamma(n+\frac{3}{2})[(n+1)k_{b}^{2}+nk_{m}^{2}]}.
\end{aligned}\tag{A5b}
\end{equation}%
Note that the spherical vector wave functions, $\vec{M}_{mn}^{s}(k_{b}r_{1}),\vec{M}_{mn}^{s(1)}(k_{b}r_{1})$ and $\vec{N}_{mn}^{s}(k_{b}r_{1}),\vec{N}_{mn}^{s(1)}(k_{b}r_{1})$, with $s=\pm$ corresponding to the TE and TM modes, respectively, can be defined as
\begin{equation}
\begin{aligned}
\vec{M}_{mn}^{\pm}(k_{b}r_{1})\approx -\dfrac{\sqrt{\pi}(k_{b}r_{1})^{n}}{2^{n+1}\Gamma(n+\frac{3}{2})}\left[\frac{mP_{n}^{m}(\kappa)}{\sin \theta_{1}}t^{\pm}_{m}(\phi_{1})\hat{\theta}\pm\dfrac{dP_{n}^{m}(\kappa)}{d\theta_{1}}t^{\mp}_{m}(\phi_{1})\hat{\phi}\right],
\end{aligned}\tag{A6a}
\end{equation}%
\begin{equation}
\begin{aligned}
\vec{M}_{mn}^{\pm(1)}(k_{b}r_{1})\approx \dfrac{i2^{n}m\Gamma(n+\frac{1}{2})}{\sqrt{\pi}(k_{b}r_{1})^{n+1}}\left[\frac{P_{n}^{m}(\kappa)}{\sin \theta_{1}}t^{\pm}_{m}(\phi_{1})\hat{\theta}+\dfrac{dP_{n}^{m}(\kappa)}{d\theta_{1}}t^{\mp}_{m}(\phi_{1})\hat{\phi}\right],
\end{aligned}\tag{A6b}
\end{equation}%

\begin{equation}
\begin{aligned}
\vec{N}_{mn}^{\pm}(k_{b}r_{1})\approx\dfrac{\sqrt{\pi}(n+1)(k_{b}r_{1})^{n-1}t^{\mp}_{m}(\phi_{1})}{2^{n+1}\Gamma(n+\frac{3}{2})}\left[ nP_{n}^{m}(\kappa)\hat{r}+\dfrac{dP_{n}^{m}(\kappa)}{d\theta_{1}}\hat{\theta}-\frac{mP_{n}^{m}(\kappa)}{\sin \theta_{1}}\dfrac{t^{\pm}_{m}(\phi_{1})}{t^{\mp}_{m}(\phi_{1})}\hat{\phi}\right],
\end{aligned}\tag{A6c}
\end{equation}%
\begin{equation}
\begin{aligned}
\vec{N}_{mn}^{\pm(1)}(k_{b}r_{1})\approx\dfrac{i2^{n}n\Gamma(n+\frac{1}{2})t^{\mp}_{m}(\phi_{1})}{\sqrt{\pi}(k_{b}r_{1})^{n+2}}\left[ -(n+1)P_{n}^{m}(\kappa)\hat{r}+\dfrac{dP_{n}^{m}(\kappa)}{d\theta_{1}}\hat{\theta}-\frac{mP_{n}^{m}(\kappa)}{\sin \theta_{1}}\dfrac{t^{\pm}_{m}(\phi_{1})}{t^{\mp}_{m}(\phi_{1})}\hat{\phi}\right],
\end{aligned}\tag{A6d}
\end{equation}%
where $t^{+}_{m}(\phi_{1})=\sin(m\phi_{1})$ and $t^{-}_{m}(\phi_{1})=\cos(m\phi_{1})$. $P_{n}^{m}(\kappa)$ is the associated Legendre polynomial with $\kappa=\cos\theta_{1}$, and $\Gamma(x)$ stands for the gamma function. Substituting all the spherical vector wave functions and the centrifugal reflection coefficients into the dyadic Green's function, one can obtain the elements of the full Green's tensor. One finds that only the three diagonal terms survive:
\begin{equation}
\begin{aligned}
G^{rr}(\vec{r}_{1},\vec{r}_{1},\omega)=A+\sum_{n=0}^{\infty}BC_{nm}\left(\frac{n+1}{n+m-1} \right)^{2} , \label{eqs2}
\end{aligned}\tag{A7a}
\end{equation}%
\begin{equation}
\begin{aligned}
&G^{\theta\theta}(\vec{r}_{1},\vec{r}_{1},\omega)=A\left(1+\xi^{2} \right)+\sum_{n=0}^{\infty}\sum_{m=0}^{n}\mathcal{D}_{nm}B\left[C_{nm}P_{n+1}^{m}(0)^{2}+C_{nm}'m^{2}P_{n}^{m}(0)^{2}\right] , \label{eqs3}
\end{aligned}\tag{A7b}
\end{equation}%
\begin{equation}
\begin{aligned}
&G^{\phi\phi}(\vec{r}_{1},\vec{r}_{1},\omega)=A\left(1+\xi^{2} \right)+\sum_{n=0}^{\infty}\sum_{m=0}^{n}\mathcal{D}_{nm}B\left[C_{nm}m^{2}P_{n}^{m}(0)^{2}+C_{nm}'(n-m+1)^{2}P_{n+1}^{m}(0)^{2} \right] , \label{eqs4}
\end{aligned}\tag{A7c}
\end{equation}%
where $A=ik_{b}/6\pi$, $\xi=k_{b}r_{1}/2$, $$B=\frac{a^{2n+1}(k_{b}^{2}-k_{m}^{2})}{4\pi (a+h_{1})^{2n+2}},$$ $$C_{nm}=\frac{n(n+m-1)^{2}}{(a+h_{1})^{2}k_{b}^{2}[(n+1)k_{b}^{2}+nk_{m}^{2}]}$$ and $$C_{nm}'=\frac{a^{2}}{2n(n+1)[4n(n+2)+3]}.$$ Sequentially, the approximate Lorentzian form of the spectral density in Eq.~(\ref{eq2}) can be obtained by substituting Eqs.~(\ref{eqs2})-(\ref{eqs4}) into Eq.~(\ref{eqs1}).
\twocolumngrid


\end{document}